\documentstyle[preprint,aps,eqsecnum]{revtex}
\begin{document}

\title
{Phonon Scattering of Composite Fermions}

\author{D.V. Khveshchenko$^1$ and M.Reizer$^2$}
\address
{$^1$ NORDITA, Blegdamsvej 17, Copenhagen DK-2100, Denmark\\
$^2$ Department of Physics, Ohio State University, Columbus,\\
OH 43210-1106}

\maketitle

\begin{abstract}
\noindent
We study the principal aspects of the interaction between acoustic
phonons and two-dimensional electrons in quantizing magnetic fields
corresponding to even denominator fractions.
Using the composite fermion approach we derive the vertex of the   
electron-phonon coupling mediated by the Chern-Simons gauge field. We 
estimate acoustic phonon contribution to electronic mobility, phonon-drag thermopower,
and hot electron energy loss rate, which all, depending on the temperature
regime, are either proportional to lower powers of $T$
than their zero field counterparts, 
or enhanced by the same numerical factor as the  
coefficient of surface acoustic wave attenuation.

\end{abstract}
\pagebreak
The discovery of gapless compressible states at even denominator fractions (EDF)
$\nu \sim 1/\Phi$, $\Phi=2,4, etc$ became a new challenge for the theory of the interacting two-dimensional electron gas (2DEG) in quantizing magnetic fields. A number of metal-like features exhibited by strongly correlated EDF electronic states   
\cite{exp} motivated the theoretical idea \cite{HLR} to describe these states 
as a new kind of Fermi liquid, which is formed by spinless fermionic quasiparticle named composite fermions
(CFs). On the mean field level the CFs, regarded as spin-polarized electrons
bound to $\Phi$ flux quanta, experience zero net field and occupy all states with momenta $k<k_{F,cf}=(4\pi n_e)^{1/2}$, where $n_e$ is the 2DEG density, inside the effective CF Fermi surface. However, as a cost of such a simplification,
the residual interactions of the CFs, as well as their interactions with charged impurities (remote
ionized donors sitting some $10^2 nm$ apart from the 2DEG) turn out to be essentially more singular than the conventional
Coulomb ones. In the framework of the Chern-Simons theory of Ref.\cite{HLR} these interactions
appear as the gauge forces mediated 
by local density fluctuations.  Conceivably, a 2D Fermi system governed by long-ranged retarded
gauge interactions could demonstrate quite unusual properties and thereby provide an example of a genuine non-Fermi liquid (NFL). 
Indeed, beyond the mean field approximation, the CF theory \cite{HLR} predicts such NFL phenomenon
as a divergence of the CF effective mass $m_{cf}^*$ already in the lowest order of perturbation theory.

Therefore, a qualitative success of the mean field CF picture in explaining the
experimentally observed Fermi liquid-like features at $\nu =1/2, 3/2$, and 3/4  
\cite{exp} caused a great deal of theoretical activity intended to reconcile these experimental observations 
 with the implications of the CF gauge theory \cite{th}. 

The present understanding of the situation is that 
the electrical current
relaxation processes, which correspond to smooth fluctuations of the ostensible CF Fermi surface,  can be safely 
described by means of the Boltzmann equation, where the singular self-energy and the Landau function terms
largely compensate each other \cite{KLW}. In particular, physical electromagnetic response functions in the low-frequency long-wavelength regime can be computed
in the framework of the random phase approximation (RPA) while ignoring the $m_{cf}^*$ divergence.
The latter, however, is expected to manifest itself in 
processes which involve rough fluctuations of the CF Fermi surface.  Those are responsible, for example, for the Subnikov-de-Haas-type oscillations of the magnetoresistivity $\Delta\rho_{xx}(B)$ in a residual field $\Delta B=B-2\pi\Phi n_e$ in the vicinity of primary EDFs, such 
as $\nu=1/2$. The theoretical 
predictions of an essentially NFL shape of the corresponding Dingle plot \cite{LMS}
is qualitatively consistent with experiment, although
the actual divergence of $m_{cf}^*(\Delta B)$ derived from the data \cite{SdH} is stronger than predicted.

Yet, even in a hydrodynamic regime described by the Boltzmann equation for   
CFs with a finite (mean field)
$m_{cf}^*\sim k_{F,cf}\epsilon_0/e^2$, one might expect that when it comes to a coupling to another subsystem, 
the CF "marginal Fermi liquid" will behave differently from the standard Coulomb-interacting 2DEG at zero field.
Such NFL deviations from the conventional behavior would then provide new  
tests for the CF theory. 

As an example of this sort, it was recently shown
\cite{DVK}  that
a combined effect of the CF gauge interactions and impurity scattering leads to the experimentally
observed non-universal $\ln T$ term in the
resistivity, which is strongly enhanced compared to its zero field universal counterpart \cite{log}. 

In the present Letter we discuss another such example provided by the electron-phonon scattering at EDF. 

In the well-studied zero-field case the electron-phonon scattering in $GaAs$ heterostructures 
 is dominated by the piesoelectric (PE) coupling at temperatures $T< 3-4K$ (see \cite{e-ph} and references
therein). Above few Kelvin the
coupling via deformation potential becomes important as well, whereas  the
coupling to optical phonons remains negligible up to $T\sim 40K$. 

In what follows we will concentrate on the range of temperatures below $1K$ where the only important coupling
is the PE one, and treat phonons as bulk acoustic modes coupled to the local
2DEG density via the vertex
$M^{PE}_{\lambda}({Q})=eh_{14}(A_{\lambda}/2\rho u_{\lambda}Q)^{1/2}$
 where ${\vec Q}=({\vec q}, q_z)$ is the 3D phonon momentum, $\rho$ is the bulk density of $GaAs$, $u_{\lambda}$
is the speed of sound with polarization $\lambda$, $h_{14}$
is the non-zero component of the piesoelectric tensor which relates a local electrostatic
potential $\phi$ to a lattice displacement $\vec u$:
$\nabla_i\phi=eh_{ijk}{\partial u_j\over \partial x_k}$, 
and the anisotropy factor $A_{\lambda}$ is given by the formulae \cite{e-ph}: 
$A_{l}={9q^2_zq^4\over 2Q^6}$, $A_{tr}={8q^4_zq^2+q^6\over 4Q^6}$.

It had been a long-standing issue of whether or not the 2DEG-3D phonon vertices get screened by the 2DEG
\cite{e-ph}, and it became customary to dress the bare 3D PE vertex with the static 2D dielectric function
$\epsilon(q)=1+H(q)(2\pi e^2\nu_F/\epsilon_0q)$, where $\epsilon_0\approx 12.9$,
$\nu_F$ is the density of states on the Fermi level, and
$H(q)=\int \int dzdz' \xi^2(z)\xi^2(z')e^{-q|z-z'|}$ is the formfactor of the quantum well given in terms of the lowest occupied subband wave function $\xi(z)\sim ze^{-z/w}$.
 
In fact,
 the PE vertex undergoes a full dynamic screening, as follows  from a systematic approach to
the electron-phonon coupling as resulting from fluctuations of the Coulomb potential associated with
lattice vibrations \cite{R}. Summing up the RPA sequence of diagrams, one arrives at the expression  
$M_{\lambda}^{PE}({Q})/\epsilon(\omega, {q})$, 
where the dynamic dielectric function
$\epsilon(\omega, {q})=1+H(q)V_{e-e}(q)\Pi_{00}(\omega, q)$
is given in terms of the 2D Coulomb $\it e-e$ interaction 
$V_{e-e}(q)=2\pi e^2/\epsilon_0q$ and  the scalar 2D polarization
$\Pi_{00}(\omega, q)$. In the presence of disorder characterized
by the elastic transport time $\tau =l/v_F$ the expression for the polarization $\Pi_{00}=\nu_F(1+i\omega/v_Fq)$ obtained in the clean limit $ql>1$ and $\omega\tau>1$ 
changes to
$\Pi_{00}(\omega, q)=\nu_F{Dq^2\over {i\omega +Dq^2}}$
at $ql<1$ and $\omega\tau<1$ (here $D=v^2_F\tau/2$ is the diffusion coefficient).  

Then, provided the ratio $u/v_F<1$ is fairly small (in what follows we will not distunguish between
longitudinal $u_l$ and transverse $u_{tr}$ sound velocities while making qualitative estimates)
the use of $\Pi_{00}(uQ, q)$ in the screened matrix element $M({Q})$ leads to the same
results as the naive static screening at all $q>u/D$ which is
equivalent to $T>T_1=\tau u^2/l^2 \sim  0.1mK$. The latter temperature is small compared to
$T_2=u/l\sim 1mK$ below which one must
use the expression for $\Pi_{00}(uQ, q)$ which contains the diffusion pole.

In turn, $T_2$ is much smaller than the Debye temperature $T_D=2uk_F\sim 10K$ 
below which both in-plane $q$ and out-of-plane $q_z$ components
of the phonon momentum $Q$ are controlled by temperature. Indeed, 
for typical electron densities $n_e\sim 10^{11}cm^{-2}$ one has $\kappa>k_F$, and
then the width of the quantum well $w\sim (\kappa n_e)^{-1/3}$ 
provides a cutoff for $q_z\sim 1/w$ which is larger than $k_F$ (in all our estimates
throughout this paper we use the typical values of the parameters from \cite{e-ph}). 
Furthermore, at these densities the PE vertex is effectively screened ($|M^{PE}_{\lambda}(Q)/\epsilon(\Omega_q,q)|^2\sim q^2/Q$) at all $T<T_D$.

{\it Phonon-limited mobility}\\
The low-$T$ the momentum relaxation due to $\it e-ph$ scattering is usually
slow compared to the impurity transport rate $1/\tau_0$, which allows one to estimate
the phonon contribution to the electronic mobility of the 2DEG by means of the standard formula \cite{e-ph}
\begin{equation}
\mu^{-1}_{e-ph}={2\pi m^*\over e}\sum_{{\vec Q},\lambda}
{|M^{PE}_{\lambda}({Q})|^2\over |\epsilon(\Omega_Q, q)|^2}
 |F(q_z)|^2  {\Omega_{Q}\over T}
N({\Omega_{Q}\over T})(1+N({\Omega_{Q}\over T}))\cos^2\theta \delta({q^2\over 2m^*}-v_Fq\cos\theta)
\end{equation}
where $\Omega_Q=uQ$ is the dispersion of phonons distributed with 
$N(x)=(e^x-1)^{-1}$ and $F(q_z)=\int dz \xi^2(z)e^{izq_z}$. In the Bloch-Gruneisen regime $T<T_D$ Eq.(1) yields $\mu^{-1}_{e-ph}\sim T^5$
which changes to a linear behavior above $T_D$ \cite{e-ph}. The theoretical prediction
of the non-linear $T$-dependence of the
phonon-limited mobility was experimentally confirmed in \cite{S}.

The above dependence, however, can only hold at $T_2<T<T_D$, whereas at lower $T$ 
the Matthiessen's rule breaks down because of the effects of quantum interference between impurity scattering and
$\it e-ph$ interactions.

This low-$T$ regime, which is hardly accessible in the zero field case, 
becomes essentially more relevant in the case of CFs, since the absolute value of the resistivity
at primary EDF is more than two orders of magnitude higher than 
at zero field \cite{exp,SdH}, and therefore
the CF transport time $\tau_{cf}$ is much  shorter than the electronic one $\tau_0$.

The low-$T$ mobility measurements similar to those of \cite{S} were recently 
performed at $\nu =1/2$ \cite{K}. The authors of Ref.\cite{K} reported a stronger temperature dependence of $\mu^{-1}_{cf-ph}$ consistent with $T^3$ at $T<T_{D,cf}={\sqrt 2}T_D$. They also supported their findings by the results of the analytic calculation presented without derivation.

Since, to the best of our knowledge, a systematic analysis of the CF-phonon problem so far
had not been made available, to facilitate our further discussion we first derive the effective CF-phonon vertex for CFs with  
Fermi momentum $k_{F,cf}={\sqrt 2}k_F$, effective mass $m^*_{cf}\sim 10m_0$, where $m_0$ is the band electron mass  
in $GaAs$, and $\tau_{cf}\sim 10^{-2}\tau_0$. 

The dynamic screening in compressible EDF states is described
in terms of a polarization tensor $\Pi_{\mu\nu}(\omega, q)$
of CFs coupled to a 2D Chern-Simons (CS) gauge field $a_{\mu}=(a_0,{\vec a})$ \cite{HLR}.

In the Coulomb gauge
$({\vec \nabla}{\vec a}=0)$ the CF polarization is a $2\times 2$ matrix
${\hat \Pi}={\it diag}(\Pi_{00}, \Pi_{\perp})$
corresponding to the scalar $a_0$ and the transverse vector $a_{\perp}$ components of the CS gauge field. 
The CS gauge propagator $U_{\mu\nu}(\omega, q)$ depends on the actual form of the {\it e-e} interaction
\begin{equation}
U^{-1}_{\mu\nu}(\omega, q)=
\pmatrix{
0 & iq/(2\pi\Phi)\cr
-iq/(2\pi\Phi) & q^2V_{e-e}(q)/(2\pi\Phi)^2}
\end{equation}
The form of the scalar CF polarization $\Pi_{00}(\omega, q)$ is similar to that of ordinary electrons, which we discussed above.
The transverse vector component is given by the formula 
$\Pi_{\perp}(\omega, q)=\chi_{cf} q^2 +i\omega\sigma_{cf}(q)$, where $\chi_{cf}\sim 1/\nu_{F,cf}$ and $\sigma_{cf}(q)$ equals $\nu_{F,cf} D_{cf}$ if $ql_{cf}<1$ and $k_{F,cf} /(2\pi q)$ otherwise.

Summing up the RPA sequence of polarization diagrams we obtain that a PE scalar potential $\phi$
generated by  lattice vibrations
induces both scalar and vector components of the gauge field acting on CFs: 
$a_{\mu}=(1-{\hat U}{\hat \Pi})^{-1}_{\mu 0}e\phi$.

Thus, the CF-phonon vertex acquires both the density- and the current-like parts
\begin{equation}
M^{cf}_{\lambda}=M^{cf,s}_{\lambda}+M^{cf,v}_{\lambda}=
{M^{PE}_{\lambda}(Q)\over \epsilon_{cf}(\omega, q)}(1+ (2i\pi\Phi)H(q)
{{\vec v}\times{\vec q}\over q^2}\Pi_{00}(\omega,q)) 
\end{equation}
where $\epsilon_{cf}=1+H\Pi_{00}V_{e-e}+H^2(2\pi\Phi/q)^2\Pi_{00}\Pi_{\perp}$. 
By virtue of the vertex (3) the CFs couple to both longitudinal ($l$-) and transverse
($tr$-) phonons unlike the isotropic 3D case
of electrons interacting via the electromagnetic gauge field, which only generates
coupling to $tr$-phonons \cite{R}.

In the diffusive regime $ql_{cf}<1$ corresponding to $T<T_{2,cf}$ the vertex
$M^{cf,s}_{\lambda}$ gets dressed by the impurity ladder 
$\sim\tau^{-1}_{cf} (D_{cf}q^2-i\omega)^{-1}$ whereas
 the vector part $M^{cf,v}_{\lambda}$ does not acquire such a pole. Even in the clean regime
$ql_{cf}>1$ the latter remains unscreened $(|M^{cf,v}_{\lambda}|^2\sim 1/Q)$ 
at all $T>T_{3,cf}=u^2k_{F,cf}\epsilon_0/e^2$. 

In the range of temperatures $T_{3,cf}<T<T_{D,cf}$ 
where the limiting 3D momentum of thermally excited phonons is controlled by $T$ and, at the same time, the dynamics of the CFs remains non-diffusive, one can use Eq.(1) and obtain  
$\mu^{-1}_{cf-ph}\sim (h_{14}^2\epsilon_0^2/e^3\rho u^4k_{F,cf})T^3$. The main contribution to
$\mu^{-1}_{cf-ph}(T)$ comes from the vector part of (3).

By contrast to Eq.(1-2) from
\cite{K} our expression for $\mu^{-1}_{cf-ph}(T)$ contains neither $m^*_{cf}$ nor  
electron band mass in $GaAs$. Well below $T_{D}$ 
the ratio $\mu_{cf-ph}/\mu_{el-ph}$ varies simply as $(T/T_D)^2$.

It is worthwhile mentioning that, in principal, it could exist a range of temperatures
$T_{2,cf}<T<T_{3,cf}$ where $|M^{cf}_{\lambda}|^2\sim q^2/Q$ and $\mu^{-1}_{cf-ph}\sim T^5$. 
However, for typical parameters $T_{3,cf}$ appears to be close to $T_{2,cf}\sim 300mK$, which is
about the lower bound of the temperature range where the reliable data  were obtained
\cite{K}.

At $T<T_{2,cf}$ the processes of small momenta transfers  contribute
to $\mu_{cf-ph}(T)$ as
\begin{equation}
\mu^{-1}_{cf-ph}={{m^*_{cf}}\over e}Im \sum_{{\vec Q},\lambda}\int {d\omega\over 2\pi} 
|M^{PE}_{\lambda}|^2|F(q_z)|^2 {D_{\lambda}(\omega, Q)}
({|M^{cf,v}|^2\over (Dq^2-i\omega)}+{v^2_{F,cf}q^2|M^{cf,s}|^2\over (Dq^2-i\omega)^3})
{\partial\over \partial\omega}[\omega\coth(\omega/2T)]
\end{equation}
where 
$D_{\lambda}(\omega, Q)=(\omega-\Omega_{\lambda}(Q)+i0)^{-1}-(\omega+\Omega_{\lambda}(Q)+i0)^{-1}$ 
is the (retarded) phonon Green function.
From this expression we obtain that within the range $T_{1,cf}<T<T_{2,cf}$ the phonon-limited CF mobility varies
as $\ln (T_{2,cf}/T)$ provided the ratio $u/v_{F,cf}$ is small enough.
At $T<T_{1,cf}$ the correction ceases to grow logarithmically and shows only a $\sim T^2$ downward deviation from its $T=0$ value 
$\mu^{-1}_{cf-ph}\sim (h^2_{14}\epsilon_0^5/\rho u k_{F,cf}e^9\tau^3_{cf})\ln (T_{2,cf}/T_{1,cf})$.

Besides the non-universal prefactor given in terms of the PE coupling, the above term is down
by an extra factor of $1/\sigma_{cf}$ compared to the $\ln T$ term resulting from interference between 
impurity scattering and the CF gauge interactions \cite{DVK}.

{\it Surface acoustic wave (SAW) attenuation}\\
It was the SAW anomaly at $\nu=1/2$ which provided the first evidence of the compressible CF states at EDFs
\cite{exp} and inspired the formulation of the CF theory \cite{HLR}. 

Following the procedure of Ref.\cite{K2} one can derive the coupling of electrons to 
SAW phonons from the bulk PE vertex:
$|M^{SAW}(q)|^2\sim \int d{z}|M^{PE}_{\lambda}(Q)|^2|F(q_z)|^2$. By contrast to the case of bulk PE phonons, the
vertex $M^{SAW}(q)$ remains finite at $q\rightarrow 0$.

The SAW attenuation is given by the imaginary part of the 2DEG density-response
function $K_{00}(\omega, q)=\Pi_{00}(\omega,q)/\epsilon(\omega,q)$ which incorporates the effects of
the dynamical screening \cite{EG}:
\begin{equation}
\Gamma_q={2\pi\over u}|M^{SAW}(q)|^2 Im K_{00}(\Omega_q,q)\sim qIm{1\over {1+i\sigma(q)/\sigma_M}}
\end{equation} 
where in the last equation we used the standard definition of the complex 
momentum-dependent conductivity $\sigma(q)=i\sigma_{M}(1-\epsilon(\Omega_q,q))$ and $\sigma_{M}={\epsilon_0u/2\pi}\sim
5\odot 10^{-7}\Omega^{-1}$. 

In the zero field case  
$\sigma_{0}(q)\sim 1/q$ at $ql_{el}>1$ whereas the momentum-dependent conductivity of a CF state at EDF $\nu=1/\Phi$ 
is inversely proportional to the CF quasiparticle conductivity \cite{HLR}: $\sigma_{\nu}(q)\approx (e^2\nu/2\pi)^2/\sigma_{cf}(q)$, which implies that $\sigma_{\nu}(q)\sim q$
at $ql_{cf}>1$. At small $q$ both physical conductivities  approach their static values, which typically
satisfy the relations:
$\sigma_{0}>>e^2/h>>\sigma_{\nu}\sim \sigma_{M}$. In this regime the SAW attenuation becomes linear in momentum:
$\Gamma_q=\gamma q$ and the coefficient $\gamma_{\nu}$ appears to be strongly
enhanced compared to its zero field counterpart $\gamma_{0}$:
\begin{equation}
\gamma_{\nu}/\gamma_{0}={\sigma_{\nu}\over {1+\sigma^2_{\nu}/\sigma^2_{M}}}{1+\sigma^2_{0}/\sigma^2_{M}\over \sigma_0}
\approx {\sigma_0\sigma_{\nu}\over {\sigma^2_M+\sigma^2_{\nu}}}>>1 
\end{equation}
in agreement with the available data on SAW propagation \cite{exp}.

{\it Phonon-drag thermopower}\\
Thermoelectric measurements probing the low-$T$ dynamics of the CF and their interactions with the phonons were recently reported \cite{Y,T}. Below $100mK$ the thermopower (TEP) $S(T)$ measured at EDF corresponding to 
$\Phi=2$ and $4$ has an approximately linear behavior
and is believed to be of the diffusion origin \cite{Y}. At higher $T$ the measured TEP
shows a non-linear dependence which was assigned to the phonon-drag contribution
$S_g$ resulting from the momentum transfer from phonons, which acquire a net flux of momentum in the presence
of a thermal gradient ${\vec \nabla} T$, to the CFs through their interaction. 

In analogy with the standard theory of the ${\it e-ph}$ interaction the thermoelectric effect can be treated in the framework
of the Boltzmann equation \cite{CB}, which yields the closed expression for the phonon-drag TEP:
\begin{equation}
S_g={\tau_{ph}\over en_eT^2}
\sum_{{\vec Q},\lambda} |M^{PE}_{\lambda}({Q})|^2 |F(q_z)|^2 
\Omega^2_{Q} {q^2\over Q^2}
N({\Omega_{Q}\over T})(1+N({\Omega_{Q}\over T}))
 Im {K_{00}(\Omega_{Q},q)}
\end{equation}
derived under the assumption that phonons equilibrate by virtue of the boundary
scattering, 
which is characterized by the relaxation time $\tau_{ph}$ proportional to the system size. 

In the case of electrons Eq.(7) yields the known result $S_{g,el}\sim T^4$ \cite{L}, which holds in the clean regime
$T>T_2$. Below $T_2$ it crosses over to  $S_{g,el}\sim  T^3$. 
According to the above discussion this can be viewed as the change from the momentum-dependent conductivity $\sigma_0(q)\sim 1/q$ to a
constant one.

On the contrary, in the case of CFs we obtain 
$S_{g,cf}\sim T^2$ at $T>T_{2,cf}$ (provided  that $T_{3,cf}<T_{2,cf}$)
and $S_{g,cf}\sim T^3$ at $T<T_{2,cf}$.
Remarkably, the exponent is higher in the dirty limit, as opposed to the
situation at zero field. This is a direct consequence
of the fact that in the clean regime the momentum-dependent
EDF conductivity $\sigma_{\nu}(q)$ grows linearly with momentum.

Although in the regime of strong disorder both the zero field and the EDF phonon-drag TEP
share the same temperature dependence $\sim T^3$, the prefactors are drastically different.
A straightforward comparison between Eq.(5) and (7) shows that the ratio $S_{g,cf}/S_{g,el}$ is equal to that
of the SAW attenuation coefficients (6).
 
The experimental data from \cite{T} show, according to the authors, an "only marginally weaker $T$-dependence for
CFs than for electrons", which suggests that the CFs are well in the disordered regime. The data, fitted in \cite{T} with
$S_{g,el}\sim T^{4\pm 0.5}$ and $S_{g,cf}\sim T^{3.5\pm 0.5}$, demonstrate a two order
of magnitude enhancement of the prefactor in the CF case. As a remark,
we note that the authors of Ref.\cite{T} analyzed their data by using
the expression for $S_g$ derived for the case of $e-ph$ coupling via deformation potential
in the clean limit \cite{CB}.

It also follows from our analysis that the above
similarity of the $T$-dependence
of the phonon-drag TEP at zero field and at primary EDF 
is not, in fact, inconsistent with the drastic difference in the corresponding phonon-limited mobilities found in \cite{K}.

{\it Hot electron energy loss rate}\\
Another informative experimental probe of the $\it e-ph$ interaction is provided by
measurements of an effective temperature of the 2DEG as a function of applied current
$T_e(I)\sim I^{2/\alpha}$, where the exponent $\alpha$ characterizes the  
energy loss rate due to phonon emission: $P\sim T_e^{\alpha}$. 

In the framework of the Boltzmann equation $P(T)$ is given by the formula 
\begin{equation}
P(T_e, T_l)={2\pi\over n_e}\sum_{{\vec Q},\lambda} |M^{PE}_{\lambda}({Q})|^2 |F(q_z)|^2 \Omega_Q(e^{\Omega_{Q}/T_l}-e^{\Omega_{Q}/T_e})
N({\Omega({Q})\over T_l})N({\Omega_{Q}\over T_e})
Im {K_{00}(\Omega_Q,q)}
\end{equation}
At zero field and $T_l<<T_e$ Eq.(8) gives in the clean limit
the standard result $P_{el}\sim T_e^5$, which is consistent with the inelastic $\it e-ph$ scattering rate $\tau^{-1}_{in,0}\sim T^3$ \cite{e-ph}. However, in the disordered regime it changes to a lower power $P_{el}\sim T_e^4$. 

On the contrary, the situation at EDF again appears to be reversed: 
the power-law dependence
$P_{cf}\sim T_e^3$, which refers to the clean regime $T_{2,cf}<T<T_{D,cf}$ and implies the inelastic $\it cf-ph$ scattering rate $\tau_{in,\nu}^{-1}\sim T$, changes to a greater power $P_{cf}\sim T_e^4$ in the dirty limit $T<T_{2,cf}$. 
Comparing the above results obtained in the regime of strong disorder we conclude that, just 
like the case of the phonon-drag TEP, the ratio $P_{cf}/P_{el}$ is given by the factor (6).

Recently the dependence $P\sim T^4_e$ was discussed in the context of transitions between adjacent quantum Hall effect plateaus (both integer and fractional) without any reference to CFs \cite{C}. 
It is tempting to identify the nearly two order of magnitude enhancement ( compared to the zero field case)
of the emission
rate, which was observed at the transition between $\nu=1/3$ and $\nu=2/5$ \cite{C}, with the CF behavior governed by the EDF state at $\nu=3/8$.  
 A systematic experimental verification of this conjecture could lend an additional support for the CF theory.
  
To summarize, we carry out a comparative analysis of the effects of the PE electron-phonon interaction 
in the 2DEG at zero magnetic field and at strong fields corresponding to 
EDF states viewed as the "CF marginal Fermi liquid".
We show that in the latter case the acoustic phonon contribution to the electronic 
mobility, the phonon-drag TEP, and the energy loss rate for hot electrons, depending on temperature, 
either contain smaller powers of $T$, or are enhanced by the numerical factor
related to the ratio between the SAW attenuation at zero field and that at EDF.
Our results reconcile the seemingly contradicting conclusions which could of been drawn from
the experimental data on phonon-limited mobilities \cite{K} and phonon-drag TEP \cite{T}.
In addition to the already existing experimental observations, we predict a strong enhancement of the hot electron energy loss rate at EDF, which is expected to be commensurate with that of the SAW attenuation.

\end{document}